\documentstyle[12pt]{article}
\newcommand{\ls}[1]
   {\dimen0=\fontdimen6\the\font 
    \lineskip=#1\dimen0
    \advance\lineskip.5\fontdimen5\the\font
    \advance\lineskip-\dimen0
    \lineskiplimit=.9\lineskip
    \baselineskip=\lineskip
    \advance\baselineskip\dimen0
    \normallineskip\lineskip
    \normallineskiplimit\lineskiplimit
    \normalbaselineskip\baselineskip
    \ignorespaces
   }

\def\d{\delta} 			\def\D{\Delta}
\def\e{\epsilon}

\def\g{\gamma} 			\def\G{\Gamma}

 		\def\L{\Lambda}
\def\m{\mu}
\def\n{\nu}
 			
\def\p{\pi}

\def\s{\sigma}

\def\ms2{m_{B^*}^{2}}
\def\fr{\frac}
\def\ba{\begin{array}}
\def\ea{\end{array}}
\def\bz{\begin{equation}}
\def\ez{\end{equation}}
\def\by{\begin{eqnarray}}
\def\ey{\end{eqnarray}}

\def\nn{\nonumber}

\newtoks\slashfraction
	\slashfraction={.13}
\def\slash#1{\setbox0\hbox{$\, #1$}
	\setbox0\hbox to \the\slashfraction\wd0{\hss \box0}/\box0}

\def \lta {\mathrel{\vcenter
     {\hbox{$<$}\nointerlineskip\hbox{$\sim$}}}}
\def \gta {\mathrel{\vcenter
     {\hbox{$>$}\nointerlineskip\hbox{$\sim$}}}}

\begin{document}


\begin{flushright}
FERMILAB-Pub-96/033-T \\
IPNO/TH 96-09 \\
\end{flushright}

\vskip 1.0truein

\begin{center}
{ \Large\bf
Dispersive Approach to Semileptonic Form-Factors  \\
in Heavy-to-Light Meson Decays}
\vskip 0.5in
{\bf Gustavo Burdman } \\
{\em Fermi National Accelerator Laboratory, Batavia, Illinois  
60510, USA.}
\vskip 0.35in
{\bf Joachim Kambor} \\
{\em  Division de Physique Th\'eorique{\footnote{Unit\'e de Recherche des 
Universit\'es Paris XI et Paris VI associ\'e au CNRS.}}, 
Institut de Physique Nucl\'eaire \\
 F--91406 Orsay Cedex, France.}
\end{center}

\vskip0.2in
\ls{1.5}
\begin{abstract}
We study the semileptonic decays of heavy mesons into light
pseudoscalars by making use of dispersion relations.
Constraints from heavy quark symmetry, chiral symmetry and
perturbative QCD are implemented into a dispersive model for the
form-factors. Large deviations from $B^*$-pole dominance 
are observed in $B\to\pi\ell\nu$. 
We discuss the model prediction for this mode
and its possible impact on the extraction of  $|V_{ub}|$.
\end{abstract}


\vskip 0.5truecm
\newpage
\pagestyle{plain}
\section{Introduction}
\label{sec:1}
Semileptonic decays of heavy hadrons are of great interest given that they 
are the cleanest way to probe the mixing between quarks of the third 
generation with those of the lighter families. The extraction of the 
Cabibbo-Kobayashi-Maskawa (CKM) matrix element 
$V_{cb}$ from the $b\to c\ell\n$ transitions is largely freed from 
theoretical uncertainties since the advent of the heavy quark symmetries
\cite{hqet}. This is the case for exclusive decays \cite{hq_exclusive} as
well as for the inclusive lepton spectrum \cite{hq_inclusive}. 
On the other hand, the  $b\to u\ell\n$ transitions involving 
the CKM element $V_{ub}$ are still plagued with large 
theoretical uncertainties. The inclusive lepton spectrum above the 
$b\to c\ell\n$ end-point, from which $|V_{ub}|/|V_{cb}|$ can be extracted,
is still the realm of models due to the breakdown of the heavy quark
operator product expansion. For the exclusive decays, the use of heavy quark
symmetry (HQS) is reduced to relating the $D$ and  $B$ 
 decays to light hadrons at fixed values 
of the recoil energy. The main shortcoming of this prediction
for $B$ decays is that 
it only covers recoil energies available in $D$ decays. This is particularly
troublesome in the $\pi$ mode, for which most of the rate might be at large
recoil energies.  It would
thus be very useful to have a full calculation of the $B\to\p\ell\n$ decay
consistent with the constraints from HQS as well as chiral
symmetry for heavy hadrons. The information from HQS is contained in the 
 scaling behavior of the form-factors with the heavy masses as well
as in the spin symmetry relations for members of the same HQS 
spin multiplet. In the decays $H\to\p\ell\n$, Chiral Perturbation Theory 
for Heavy Hadrons (CPTHH) \cite{wise,bd}
promotes nearest singularity (pole)  dominance to the leading contribution
in the chiral and heavy quark expansions when the pion is soft. 
The validity
of nearest pole dominance in the soft pion limit is not a surprising 
result given the proximity of the pole to the physical region where
the pion recoil is small. However, there is no reason to believe that 
this is also true at higher recoil momenta. Deviations from 
the pure pole behavior at pion recoil energies above $\approx 1$GeV
in $B\to\p\ell\n$ result in large modifications in the branching
ratio. Conversely, the physical region in the case of the $D$ modes
is confined to be closer to the pole and therefore one expects
the  approximation of the form-factors by the single pole to be a rather
good one over a large fraction of phase space. In this paper, we 
construct a model which reflects pole dominance
at low pion energies but at the same time includes other effects
that are potentially important, mostly in $B$ decays. These 
will include the effect of resonances other than the $B^*$-pole as 
well as the continuum. A natural model emerges from the 
dispersion relations for the form-factors \cite{diff_bgl}.
 Their properties in the 
physical region are determined not only by the isolated poles
but also by  the singularities  above the $(m_H+m_\p)^2$ threshold. 
 The contributions from resonances above threshold can be parametrized 
in a way compatible with HQS. This allows us to fix the 
parameters of the model in $D$ decays and have a prediction
for {\em all the available phase space } in 
$B\to\pi\ell\n$.

Recently CLEO has  observed
this decay for the first time and measured its branching ratio
 \cite{cleo_bpi}. 
It is therefore imperative to realistically assess the theoretical
uncertainties associated with this mode and their impact on the extraction
of $|V_{ub}|$. The model we present here is an attempt to address this 
issue. We expect the contributions to the dispersion relations
governing these transitions to be highly constrained by chiral symmetry, 
HQS and perturbative QCD, to the point of having a complete
picture of the physics involved and very little freedom. 
In the next section we discuss the various contributions to the 
form-factors in $H\to\pi\ell\nu$, with $H=(D,B)$,  in the language of
dispersion relations.
 In Section~3 we derive constraints from chiral 
symmetry, HQS and perturbative QCD which will highly determine 
those contributions. In Section~4 we present a model that naturally
emerges from all the theoretical constraints, discuss its 
predictions and compare them with other calculations. 
We summarize our results and conclude in Section~5.

\section{Dispersion Relations}

The hadronic matrix element for the $B^0\to \pi^-\ell^+\n$ transition can be 
written as 
\bz
\langle \p({\bf p_\p}) |\bar{u} \g_\m b | H({\bf P})
\rangle = f_+(q^2) (P+p_\p)_\m + f_-(q^2) (P-p_\p)_\m \label{ff_def}
\ez
where $q^2=(P-p_\p)^2$ is the momentum transferred to the leptons. 
In the approximation where the leptons are massless, only 
the  form-factor $f_+(q^2)$ enters
the partial rate. This form-factor obeys a dispersion relation 
of the form
\bz
f_+(t)=\fr{-\g}{(m_{H^*}^{2}-t)}+\frac{1}{\pi}\int_{s_0}^{\infty}
\frac{Im\left[f_+(s)\right]ds}{(s-t-i\epsilon)} \label{un_dr}
\ez
where $s_0=(m_H+m_\pi)^2$. The isolated pole at $m_{H^*}<s_0$ is actually 
present in the $B$ meson case, whereas for the $D\to \p$ transition
the $D^*$ pole is located almost exactly at threshold. 
There are no anomalous thresholds in the $H\to\pi\ell\nu$
transitions, although there might be in $H\to\rho\ell\nu$. 
Contributing to the imaginary part in (\ref{un_dr}) are all possible
 intermediate states that couple to $H\pi$ and are annihilated by the weak 
vertex, as shown schematically in Fig.~1. 
These include the multiparticle continuum as well as resonances. The latter
 must be {\em radially} excited $J^P=1^-$ states in order to contribute to 
$f_+(t)$, and are expected to be located at about $\approx 1$ GeV above the
 ground state \cite{ehq}.
The first contribution appearing above $s_0$
 corresponds to the $H^{(*)}\pi$ continuum.  
Other contributions to the continuum include states with $H^{(*)}$ mesons and
 various light mesons (e.g. multipion states). We will neglect them
 at these values of $s$ because they involve higher 
order terms in the chiral expansion. 
Finally, we consider the contributions of orbitally excited $H$ mesons with
 one pion. In principle these could be important given that the $L=1$, 
P-wave
 states are only about $\approx 500$ MeV above the ground state \cite{ehq2}.
 The lightest $L=1$ states correspond to $(0^+,1^+)$ and $(1^+,2^+)$ doublets. 
The second doublet, however, does not couple to the ground state to leading
 order in CPTHH
 \cite{falk_luke}. 
The doublet $(0^+,1^+)$ does couple to the ground state doublet, but the
 vertex $H\pi\to (0^+,1^+)\pi$  vanishes to leading order. 
Therefore, not far from threshold, the continuum contribution to the 
dispersive integral (\ref{un_dr}) can be approximated by the $H^{(*)}\pi$ 
intermediate states. We will estimate this contribution in 
CPTHH in Section~3.1.   

The contributions from the radial excitations of the $H^*$ dominate the 
imaginary part at intermediate values of $s$. Separating them from the 
continuum one can express (\ref{un_dr}) as 
\bz
f_+(t)=\fr{-\g}{(m_{H^*}^{2}-t)}+
\frac{1}{\pi}\int_{s_0}^{\L^2}\frac{Im\left[f^{\rm cont.}_+(s)\right]ds}
{(s-t-i\epsilon)} + \sum_{i}a_i {\cal R}_i(t) . 
\label{sep_dr}
\ez
where the cutoff in the continuum integral defines the 
maximum center of mass energy in $H\p$ scattering for which the main 
contribution comes from the continuum. This also corresponds to the beginning
of the resonance-dominated region. Typically, $\L$ defines an 
energy about $0.7 {\rm ~to~} 1$ GeV above threshold. Therefore, the use 
of CPTHH to compute the continuum contribution to the dispersive integral
is justified. 

In the narrow width approximation 
\bz
{\cal R}_i(t)\equiv \fr{1}{M^{2}_{i}-t} \label{nwa_r}
\ez
whereas if finite width effects are taken into account these functions
take the form
\bz
{\cal R}_i(t)\equiv \fr{1}{\p}\left(\fr{\p}{2}-\arctan{\fr{s_0-M^{2}_i}{
M_i\G_i}}\right) \fr{(M_{i}^{2}-t)}{(M_{i}^{2}-t)^2+M_{i}^{2}\G_{i}^{2}}
\label{fw_r}
\ez
with $M_i$ the mass of the $i$-th radial excitation. 
In deriving (\ref{fw_r}) the widths $\G_i$ were assumed to be constant. 
The $a_i$'s are the couplings analogous to $\g$ of the nearest resonance
$H^*$, the residues at the poles. They involve the strong coupling 
between the {\em i}-th resonance and $H\p$, as well as the decay 
constant of the resonance. 

The physical region for the decay $H\to\p\ell\n$ is given by the interval
$t=(0,t_{\rm max})$, with $t_{\rm max}=(m_H-m_{\p})^2$. Thus the 
$H^*$-pole, the first term in (\ref{sep_dr}), corresponds to the singularity
 closest to this region
and will be the dominant contribution for values of $t$ close to 
$t_{\rm max}$. However, the question of how good this approximation is 
in each case is not a simple one. For instance, neglecting the continuum
contribution, it is likely that the $H^*$-pole approximation will be a good
one as long as the three-momentum of the recoiling pion, $p_{\p}$, 
is not larger than $\D_1\equiv M_1-m_{H^{(*)}}$, the gap between the 
ground state and the first radially excited state. This intuitive picture
suggests that the $H^*$-pole term in (\ref{sep_dr}) should be a reasonable 
approximation to $f_+(t)$ for $D$ semileptonic decays, given that almost
all its phase-space falls in this region. 
However, this is certainly not the case for $B\to\p\ell\n$. Although the 
spacings between resonances and the ground state are independent of the 
heavy quark mass to a very good approximation, now the recoil momentum 
of the pion can be as large as $p_{\p}^{\rm max}\approx m_B/2$. 
As $p_{\p}$ increases and we move away from the $H^*$-pole, the relative
influence of the higher resonances grows. These deviations 
from the pure $H^*$-pole behavior at large values of $p_{\p}$
 are particularly
important given that the pion momentum distribution
goes as $p_{\p}^3$. 
As we will see below, large changes in the total rate and the shape of 
the $t$ distribution occur when the corrections to the $H^*$-pole
behavior at large $p_{\p}$ are taken into account. 

In what follows, we will consider the theoretical constraints 
that can be imposed on $f_+(t)$. These constitute the basis for 
a model calculation of the $B\to\p$ transition form-factor, 
which will incorporate all of these constraints. 

\section{Theoretical Constraints}
Although the form-factor $f_+(t)$ is not calculable
in perturbation theory, it must satisfy several constraints. 
These result from the application of HQS, chiral symmetry 
and the asymptotic behavior imposed by perturbative QCD 
for exclusive processes. 

\subsection{Chiral Symmetry}
Chiral Perturbation Theory for Heavy Hadrons (CPTHH) provides 
a formal framework for the approximation that 
only keeps the first term in (\ref{un_dr}) \cite{wise,bd}.
In the effective theory that couples heavy hadrons to 
goldstone bosons respecting HQS and chiral symmetry, the 
heavy meson fields are represented by the $4\times 4$ matrices
\bz
{\cal H}=\fr{(1+\slash v)}{2}\left\{\slash H^*-H\g_5\right\} \label{grst}
\ez
where $H_\mu$ and $H$ are the $1^-$ and $0^-$ ground state fileds
respectively, and $v_{\mu}$ is the heavy meson four-velocity. 
The goldstone bosons enter through
\bz
\xi=\exp{(i\pi(x)_aT_a/f)} \label{xi}
\ez
with $T_a$, $(a=1,\ldots,N^2-1)$ the $SU(N)$ generators and $\pi_a(x)$  the
 goldstone boson fields. To leading order $f=f_\pi$, the pion decay constant.
The leading order Lagrangian, invariant under HQS and chiral symmetry, 
is given by \cite{wise,bd}
\bz
{\cal L}_{\rm eff.}= iTr\left[\bar{{\cal H}}v.D {\cal H}\right] +
g Tr\left[ \bar{{\cal H}}\slash A {\cal H}\gamma_5\right] \label{l_strong} .
\ez
The coupling constant $g$ is independent of the heavy mass and the axial-vector
 field is defined by
\bz
A_\mu=\frac{-i}{2}\left(\xi^\dagger\partial_\mu\xi -\xi\partial_\mu\xi^\dagger
\right) \label{a_mu} .
\ez 
Requiring that the weak current transforms as a left-handed
doublet implies 
\bz
J_{\rm weak}^{\mu}= -i\frac{\sqrt{m_H}f_H}{2}
Tr\left[\bar{{\cal H}}\xi^{\dagger}\gamma^{\mu}(1-\gamma_5)\right] 
\label{w_current}
\ez
with $f_H$ the heavy pseudoscalar decay constant. 
This completes the description to leading order 
 both in $1/m_H$ and $1/\L_{\chi}$, where 
$\L_{\chi}$ is the scale of chiral symmetry breaking.
The $H\to\pi\ell\nu$ transition receives a direct contribution
from (\ref{w_current}), as well as an $H^*$-pole term resulting
from the $H^* H\pi$ interaction governed by the coupling $g$ in 
(\ref{l_strong}) followed by the $H^*$-vacuum transition
governed by (\ref{w_current}). The latter dominates the former
in the $1/m_H$ expansion. Thus, for soft pion momentum, the 
transition form-factor can be approximated by
\bz
f_+(t)\simeq -\fr{g\;\; f_H\;\; m_{H}^2/f_{\pi}}{(m_{H}^2-t)} .
\label{chi_pole}
\ez
Therefore, CPTHH tells us that to leading order the form-factor
is approximated by the first term in the dispersion relation
(\ref{un_dr}) with 
\bz
\g=\fr{g\, f_H\, m_{H}^{2}}{f_{\p}} \label{chi_gam} .
\ez
In this way CPTHH provides an approximate normalization of $f_+(t)$
at low pion recoil momentum, $|\vec{p_\p}|\lta 1$GeV.

On the other hand, CPTHH can be used to compute the 
continuum contribution in (\ref{sep_dr}) for values of the integrand
close to threshold. At this values of $s$ unitarity implies
\bz
Im[f_{+}^{\rm cont.}(s)]=\s(s)\, T^\dagger (s)\, f_+(s)
\label{uni_con}
\ez
where the threshold factor is 
\bz
\s(s)\equiv \left( 1-\fr{(m_H^2-m_\p)^2}{s}\right)^{1/2}
\left( 1-\fr{(m_H^2+m_\p)^2}{s}\right)^{1/2} \label{sigma}
\ez
and $T(s)$ is the $H\p\to H\p$ scattering amplitude 
projected onto the $J=1$ partial wave. 
We neglect the contribution of the $H^*\pi$ intermediate
state which is suppressed by a factor of $p_{\pi}/m_H$ relative
to $H\pi$. 
Therefore, above $s_0$ but 
below the resonance 
region in the cut, the phase of $f_+(s)$ is given by the 
$H\p\to H\p$ scattering phase-shift
\bz
\sin{\delta_+}(s)=\sigma(s)\; |T(s)| \label{del_def} .
\ez
The computation from ({\ref{l_strong}) is straightforward. After
projecting to the correct partial wave and isospin channels
the phase, to leading order, is given by
\bz
\sin{\delta_+}(s)\simeq -\fr{1}{24\pi}\left(\fr{g}{f_\pi}\right)^2
p_{\pi}^{3}\left(\fr{3}{v.p_{\pi}-\Delta}+\fr{1}{v.p_{\pi}+\Delta}
\right) \label{del_chi}
\ez
where $\Delta\equiv m_{H^*}-m_H$ is the mass splitting within the 
ground state and 
\bz
v.p_{\pi}=\fr{s-m_{H}^2-m_{\pi}^2}{2\;\;m_{H}}
\label{v_pi}
\ez 
is the pion energy in the $H$ rest frame. 
The dominant effect, once again, comes from the $H^*$-pole exchange, 
both in the $s$ and $t$ channels. These are governed by the
same coupling $g$ entering in the $H^*H\p$ vertex in (\ref{l_strong}), 
and therefore do not introduce any new parameter.
The phase in (\ref{del_chi}) together with the leading order expression 
for $|f_+(s)|$ yields $Im[f_{+}^{\rm cont.}](s)$, in the threshold region. 
The corresponding dispersive part provides, in a next-to-leading singularity
approximation, the first correction to the $H^*$-pole behavior.
However, as we will see below, this is not the
most important modification to pole dominance coming from the cut.
The presence of 
radially excited states, coupling to both the ground state and the weak
current turns out to be a more significant correction. 

\subsection{Heavy Quark Symmetry}
Although HQS  is an ingredient of the CPTHH formulation, 
the applicability of the flavor symmetry is very limited in practice, 
as it was mentioned above, if the scaling of $f_+(t)$ with $m_H$
is used to relate $D\to\p\ell\n$ to $B\to\p\ell\n$. 
However, we will show here that the application of HQS to the 
dispersion relation of (\ref{sep_dr}) does not suffer from  such
limitations. The first two terms in (\ref{sep_dr}) already have 
a defined behavior with $m_H$ built in CPTHH. 
On the other hand, the residues $a_i$ governing the resonant contributions
scale as $m_H^{3/2}$,  the same way $\g$ does. This allows us to write
\bz
\fr{a_i}{\g} =\fr{g_i}{g}\, r_i \label{ais_gam}
\ez
where $g_i$ is a dimensionless constant characterizing the coupling of the 
radially excited resonance $H_i$ to $H^{(*)}\p$, and $r_i$ is essentially
the ratio of the mass independent decay constant of $H_i$ to that 
of the ground state, $f_H$. 
On the other hand, the spacing among resonances and also the 
gap to the ground state is independent of $m_H$, to leading order
in the heavy mass. Therefore, if we knew the 
$a_i$'s and the masses of the resonances
for the charm system, we would know $f_+(t)$ for $B\to\p\ell\n$
{\em in the whole physical region}.  
This is an important departure from the application
of the flavor HQS to semileptonic decays. The HQS is applied to the 
resonances in the cut, which have excitation energies independent
of the heavy mass as long as both the $c$ and the $b$ quark are 
considered to be sufficiently heavy. 
Of course, we do not know a priori the values of the $a_i$'s and 
the spacings among  resonances and the ground state. 
The latter can be calculated in 
 potential
models that  have been successful in predicting the spectrum of the
{\em orbitally} excited heavy mesons \cite{ehq2} and are expected
to yield a good approximation also for the radial excitations. 
Regarding the couplings $a_i$, if the sum in (\ref{sep_dr}) has a small
number of terms, we will show below that the asymptotic behavior
of $f_+(t)$  plus the $D$ decay data can be used to fix their values, 
resulting in a fully predictive model for the $B$ decay. 
This procedure assumes that the effect of the heaviest resonances
can be absorbed in the values of the lighter radial excitations
in such a way that the ``effective"  couplings
still obey the heavy mass scaling given above. 
For instance, truncating the sum over resonances at $i=2$,  the effect of the
 heavier resonances can be absorbed, for instance, by $a_2$, defining
\bz
a_{2}^{\rm eff.}\equiv a_2 + \sum_{i=3}\, 
a_i\left(1-\fr{\d_i}{v.p_\p+\D_2}
\right)\, \left(1-\fr{\d_i}{2m_H}\right)
\label{a2_eff}
\ez
with $\D_2\equiv M_2-m_H$ and $\d_i\equiv M_i-M_2$. As long as 
the spacing between successive resonances becomes smaller, one can neglect
$\d_i/(2m_H)$ corrections. This implies that $a_2^{\rm eff.}$ has the 
same $m_H$ dependence as $a_2$ so the $m_H$ scaling is still 
valid for the truncated case. 
The interpretation of $a_2^{\rm eff.}$ is not straightforward. In 
particular, there is no clear correspondence  of this quantity  with
the product of the coupling $g_2$ times the decay constant of the  
resonance $H_2$. This, however,  is not a problem as we will fit the
effective parameters to data. 
Before doing so, there is a last theoretical constraint we can impose
on $f_+(t)$, regarding its behavior at large values of $|t|$.

\subsection{Asymptotic Behavior}
The behavior of the form-factor $f_+(t)$ for very large values of $|t|$
can be estimated reliably in perturbative QCD for exclusive 
processes (pQCD) \cite{pqcd}. In this approach the hadronic 
 matrix element is described by the hard scattering transition amplitude
folded into an overlap integral between the two hadron state wave-functions.
To leading order, the hard scattering amplitude is approximated by 
the one-gluon exchange diagrams. The gluon momentum satisfies
\bz
Q^2=(1-x)^2m_{H}^{2}+(1-y)^2m_{\p}^{2}\pm 2(1-x)(1-y)P\cdot p_{\p}
\label{glu_mom}
\ez
where $x$ and $y$ are the momentum fractions of the non-spectator quarks
in the initial and final hadron, respectively. 
The positive sign in the last term in (\ref{glu_mom}) corresponds to
the s-channel process $\ell\n\to H\p$, whereas the negative signs 
corresponds to the t-channel, e.g. $\ell H\to \n\p$ as well as to the 
decay $H\to\p\ell\n$. In order for pQCD to be safely applicable we 
need $Q^2\gg 1 {\rm GeV}^2$.  However, the wave-function of a meson 
containing a heavy quark peaks at $x\simeq (1-\epsilon)$, with 
\bz
\epsilon\simeq {\cal O}\left( \fr{\L_{\rm QCD}}{m_Q}\right)
\label{epsilon}
\ez
This implies that in the physical region for the decay $H\to\p\ell\n$
the gluon momentum is $Q^2\lta 1{\rm GeV}^2$, with the exception of 
a negligible large-$Q^2$ tail of the wave-function. This casts a serious 
shadow over the applicability of the one-gluon exchange approximation 
in computing $f_+(t)$ in the physical region for the semileptonic
decay, signaling possible large corrections. However, outside the 
physical region and for large enough values of $|t|$, the condition
$Q^2\gg 1{\rm GeV}^2$ is satisfied. 
In these two regions, for $t\ll 0$ and for $t$ much larger than the 
typical mass of heavy resonances, pQCD should yield a 
very good  approximation
to the form-factor. The known asymptotic behavior of 
$f_+(t)$ constitutes an important constraint to be 
satisfied by any calculation. We rewrite the dispersion relation
as
\by
f_+(t)&=&-\fr{\g}{(m_{H^*}^2-t)} + \fr{1}{\p}\int_{s_0}^{\L^2}
\fr{Im[f_{+}^{\rm cont.}(s)] ds}{s-t-i\e} 
+ \fr{1}{\p}\int_{\L^2}^{\L'^2}\fr{Im[f_{+}(s)] ds}{s-t-i\e} \nn\\
&+& \fr{1}{\p}\int_{\L'^2}^{\infty}\fr{Im[f_{+}(s)] ds}{s-t-i\e}
\label{dr_asy}
\ey
The second  term in ({\ref{dr_asy}) contains the continuum contribution, 
whereas the third one accounts for the region dominated by resonances, 
for $\L^2 < s < \L'^2$. The last term in (\ref{dr_asy}) can be calculated 
perturbatively provided $\L'$ is sufficiently large.
Its contribution is small for values of $t$ in 
the physical region. At very large $|t|$, for 
instance for $t\ll 0$, the form-factor can also be calculated perturbatively. 
Thus, the asymptotic behavior of (\ref{dr_asy}) gives 
\by
f_+(t)&\longrightarrow &\fr{-1}{t}\left\{-\g +\fr{1}{\p}\int_{s_0}^{\L^2}
Im[f_{+}^{\rm cont.}(s)]ds +\fr{1}{\p}\int_{\L^2}^{\L'^2}
Im[f_{+}(s)]ds +p_1(t,\L') \right\} \nn \\
 &\longrightarrow &f_+^{\rm pQCD}(t)
\label{fp_asy}
\ey
where the last term in the brackets in (\ref{fp_asy}) is the leading 
asymptotic contribution from the last term in (\ref{dr_asy}). 
The  non-perturbative
contributions in  (\ref{fp_asy}) are, {\em individually}, much larger
than $t\times f_+^{\rm pQCD}(t)$. Therefore, because $f_+^{\rm pQCD}(t)$
is a reliable approximation to the form-factor for $t\to -\infty$, there
must be large cancellations among the non-perturbative contributions. 
This leads to a convergence condition or ``sum rule" of the form:
\bz
\g -\fr{1}{\p}\int_{s_0}^{\L^2}Im[f_{+}^{\rm cont.}(s)]ds 
-\fr{1}{\p}\int_{\L^2}^{\L'^2}Im[f_{+}(s)]ds \simeq 0 
\label{con_sr}
\ez
where the equality corresponds to 
$f_+^{\rm pQCD}(t)=-p_1(t,\Lambda ')/t$. 
This sum rule translates our knowledge of the asymptotic behavior
of $f_+(t)$ as a constraint on the non-perturbative contributions, 
which in turn dominate $f_+(t)$ in the physical region. 
In order to actually implement this constraint, we can
write the integral over the resonant region as a sum over the 
individual resonances.  In the narrow width 
approximation, we have
\bz
\fr{1}{\p}\int_{\L^2}^{\L'^2} \fr{Im[f_+(s)]ds}{s-t-i\e} = \sum_{i=1}
\fr{a_i}{M_{i}^2-t} \label{ai_def}
\ez
where the $a_i$'s are defined in Section~2. In this way the convergence
condition now reads
\bz
\g - c - \sum_{i=1}a_i \simeq 0 
\label{con_nwa}
\ez
where we defined
\bz
c\equiv \fr{1}{\p}\int_{s_0}^{\L^2} Im[f_+^{\rm cont.}(s)]ds \label{def_c}
\ez
Finite width effects are taken into account by the replacement 
$a_i \to a_i I_i$, with the $I_i$'s defined by
\bz
I_i\equiv \fr{1}{\p} \left( \fr{\p}{2} - \arctan{\fr{s_0-M^{2}_i}{
M_i\G_i}}\right) .
\label{con_fwa}
\ez
The convergence relation of (\ref{con_nwa}) is a very binding
constraint for model building. In the next section we will explore
a specific model for the resonant contribution in order 
to understand the effect of (\ref{con_nwa}) on the behavior
of $f_+(t)$ in the physical region.

\section{Constrained Dispersive Model of  $f_+(t)$} 
In order to implement the various theoretical constraints from 
the previous section into a calculation of $f_+(t)$ we need to 
specify the resonant contributions. First, we must truncate the 
sum over resonances in (\ref{sep_dr}). As the radial excitations of the 
H* become heavier, they are less relevant to $f_+(t)$.  On the one hand, 
heavier resonances contribute with a smaller value of ${\cal R}_i$ even 
in the narrow width approximation. Furthermore, as finite widths are
considered, the contributions of heavier and thus broader excitations
are additionally suppressed, as can be seen in (\ref{fw_r}). On the other
hand, the couplings of the excitations to the ground state, $g_i$, 
are constrained to obey an  Adler-Weisberger sum rule \cite{sum_rule}
\bz
1\gta g^2 + g_{1}^2 +g_{2}^2+ ...
\label{sum_rule}
\ez
where $g_i$ is the coupling the the {\em i}-th radial excitation 
to $H\p$, and 
additional terms, e.g. from orbitally excited states, 
are not shown. This implies that one cannot add a large number 
of resonances in the cut with large couplings to the ground 
state. This, together with the mass and width suppression, 
shows that the truncation of the sum over resonances is a 
controlled approximation. 

In what follows, we will study a 
Constrained Dispersive Model (CDM) 
where only the first two terms in 
the sum over resonances are kept. This is partly motivated 
by the fact that only two $1^-$ radially excited states are known 
in the light-quark sector. On the other hand, the ``minimal" choice of 
keeping only one term will turn out to be incompatible with the data on
exclusive  charm semileptonic decays and the convergence condition
(\ref{con_nwa}), as we will see below. 

The other necessary ingredient to specify the model is the knowledge
of the spectrum of radial excitations. These resonances ((2S) and 
(3S) excitations of $D^*$ and $B^*$) have not been observed in the 
$D$ or $B$ systems. We will then rely on potential model calculations 
of their 
masses \cite{ehq}. These models have been very successful in predicting the 
masses of orbitally excited states, and therefore we are confident that 
the position of the radial excitations in the cut does not introduce a 
sizeable uncertainty. The resulting spectrum explicitly shows that the
spacings among the 1S, 2S and 3S states are, to leading order, independent
of the heavy quark mass and therefore constitute a property of the 
light degrees of freedom.  We take the spectrum of radial excitations
to be \cite{ehq}
\bz
\ba{cc}
M_{1}^{D}=2.7 {\rm ~GeV}\;\;\; ; & M_{2}^{D}=3.3 {\rm ~GeV} \\
M_{1}^{B}=6.1 {\rm ~GeV}\;\;\; ; & M_{2}^{B}=6.6 {\rm ~GeV} 
\ea
\label{mass_spec}
\ez
where the subindex $1$ corresponds to the $2$S excitation of the 
$H^*$, etc. 
In this model, the convergence condition (\ref{con_nwa}) now reads
\bz
\g-c-a_1-a_2\simeq 0 
\label{cr_mod}
\ez
This condition, together with eqn.~(\ref{chi_gam}) for $\g$ and the 
CPTHH calculation for the continuum term $c$ defined in (\ref{def_c}),
leaves only one free parameter in the model. This parameter can be fixed by
fitting to the observed $D^0\to\p^- e^+\n_e$ branching ratio. 
Given that $a_1/\g$ and $a_2/\g$ are independent of the heavy
quark mass, this procedure results in a prediction for 
$B^0\to\p^-\ell^+\n$. The result not only has the correct scaling 
of $f_+(t)$ with the heavy meson mass but also implements the 
HQS properties of the resonances in the cut, namely the scaling of 
the $a_i$'s with $m_H$. 

The form-factor emerging in this picture has the form
\by
f_+(t)&=&\fr{-\g}{m_{H^*}^2-t}\left(\fr{M_{2}^2-m_{H^*}^2}{M_{2}^2-t}\right)
+\fr{a_1 (M_{2}^2-M_{1}^2)}{(M_{1}^2-t)(M_{2}^2-t)} \nn \\
& &+\fr{1}{(M_{2}^2-t)}\fr{1}{\p}\int_{s_0}^{\L^2} \fr{(M_{2}^2-s)
Im[f_{+}^{\rm cont.}(s)]}{s-t-i\e} ds
\label{fp_nwa}
\ey
where we made use of the narrow width approximation. The cutoff in the 
integral in (\ref{fp_nwa}) is given by $\L\simeq (m_{H}+0.7{\rm GeV})$, 
which corresponds to the maximum center-of-mass energy in $H\p$ scattering
for which CPTHH can be used to compute the phase of $f_+(s)$, 
and at the same time coincides with the beginning of the resonance 
dominated region in the cut.

In order to fit (\ref{fp_nwa}) to the $D^0\to\pi^-e^+\n_e$ data 
we need values for $f_D$ and $g$ entering in the expression 
(\ref{chi_gam}) for 
$\g$. Currently there is only an upper limit on the decay constant
from $D\to\mu\nu$ , $f_D<0.310$~GeV \cite{pdg}. On the other hand, 
the CLEO collaboration measurement of  $D_s\to\mu\n$ gives
$f_{D_s}=(0.284\pm0.030\pm0.030\pm 0.016)~$GeV \cite{fd_exp}.
 We combine this with the predictions
from lattice calculations \cite{latt}
for the ratio of decay constants, to obtain 
the value $f_D=0.24$~GeV, to be used in the fit. The $H^*\; H\p$
coupling gets an upper bound from the upper limit on the $D^*$ lifetime
 \cite{accmor} plus the $D^*\to D\p$ branching fraction \cite{pdg}. 
It is also possible to derive a lower limit on $g$ from $D^*\to D\g$ 
\cite{boy_ros}. These two combined give $0.3 < g < 0.7$. We take 
$g=0.50$ for some of our numerical estimates. 
Finally, the prediction for $B^0\to\pi^-\ell^+\n$ will depend on 
the $B$ meson decay constant $f_B$. Lacking experimental information
on this quantity, we make use of the scaling with heavy meson masses
and, from the value for $f_D$, we obtain $f_B=0.15$~GeV. In any event the 
values of $f_D$, $f_B$ and $g$ are external input parameters and not 
model parameters. Eventually, they will be experimentally
determined. 

We are now ready to fit (\ref{fp_nwa}) to $D^0\to\p^- e^+\n_e$ and 
fix the value of $a_1/\g$. A recent CLEO measurement \cite{cleo_dp}, 
combined with the value in \cite{pdg} for $D^0\to K^- e^+\n_e$, 
gives $Br^{\rm CLEO}(D^0\to\p^- e^+\n_e=(3.90\pm 1.5)\times 10^{-3}$. 
We do not make use of the more precisely 
measured  $D^0\to K^-\ell^+\nu$ mode due to the presence of 
potentially large $SU(3)$ breaking effects, which we do not take into
consideration. 
 Thus, at this point, the largest uncertainty in the model prediction
comes
from the experimental uncertainty in the $D^0\to\p^- e^+\n_e$
branching ratio. It is expected that this will be known to within $5\%$
in the near future \cite{will_johns}.  

With this choice of external parameters, the form-factor of (\ref{fp_nwa})
is shown in the solid line of Fig.~2 for the $B^0\to\pi^-\ell^+\nu$
mode. 
Also shown for comparison, are
the first term in the dispersion relations (\ref{un_dr}) (dashed-line),
corresponding to the $B^*$-pole with $\g$ given by (\ref{chi_gam}) (chiral
$B^*$-pole), as well as the prediction from the BSW model (dotted line)
 \cite{bsw}. 
The pion
momentum distributions for these cases are shown in Fig.~3. 
As expected, the chiral $B^*$-pole is a good approximation to the full 
form-factor up to pion momenta of ${\cal O}(1)$~GeV. Above this 
momentum, large deviations from the $B^*$-pole behavior are observed. 
On the other hand, with the choice of parameters made, the model 
interpolates between the soft momentum region, predicted 
by CPTHH, and the BSW calculation of the form-factor at $t=0$. This 
type of procedure was first suggested in \cite{bd_pole}. 
After all, the relativistic quark model calculation of 
$f_+(0)$ in the BSW model constitutes the most reliable prediction
of quark models for  charmless decays of $B$ mesons.  Thus, one
alternative within our model to the need of knowing the external
parameters, is to normalize the prediction in $B\to\pi\ell\nu$ to 
agree with the prediction from \cite{bsw} at $t=0$. As seen in Fig.~3, 
choosing the central values of all the external parameters 
provides us with a good extrapolation. 
On the other hand, the ISGW2 model \cite{isgw2} is a modification
of the non-relativistic quark model of \cite{isgw} which has a
harder pion momentum spectrum than the original calculation in order
to give a better fit to the pion electromagnetic form-factor $F_{\p}(Q^2)$. 
Thus the shape of the spectrum in the ISGW2 model is 
a parametrization of the soft physics governing $F_{\pi}$. The 
{\em shape} of the  resulting $f_+(t)$
resembles the one obtained in our calculation and is 
yet another reason to believe that the convergence relation
captures the correct physics at large values of $t$.
In the future, the 
precise measurement of $g$, $f_D$ and $f_B$ will provide an independent 
prediction whithin our model
and will help us understand the relation between this simple
dispersive approach and quark models.  

We now address the possible sources of corrections to this calculation
of the form-factor. 
The widths of the resonances are expected to be rather large.  
When finite width effects are taken into account, the form-factor
is that given by (\ref{sep_dr}) keeping  only the first two terms in 
the sum over resonances, with ${\cal R}_{1,2}$ given by (\ref{fw_r})
and the convergence relation
\bz
\gamma - c -a_1 I_1 -a_2 I_2 \simeq 0 \label{cr_fw}, 
\ez
where $I_1$ and $I_2$ are defined by (\ref{con_fwa}). 
The widths $\G_1$ and $\G_2$ are, in principle, not fixed. However, they can 
be estimated by calculating the partial widths to the ground state as well 
as to orbital excitations. To leading 
order in CPTHH, these interactions are governed by 
\bz
{\cal L}=\sum_{i=1}^{2} g_i Tr[\bar{\cal E}_i \slash A 
{\cal H} \gamma_5] 
+ \sum_{i=1}^{2} f_i Tr[\bar{\cal E}_i \slash A {\cal S} \gamma_5]
\label{rad_lag}
\ez
where ${\cal H}$ is defined in (\ref{grst}) and represents the ground state
doublet, ${\cal E}_i$ represent the two radial excitation doublets and 
have the same form as ${\cal H}$, and the $(0^+,1^+)$ P-wave doublet is 
given by \cite{falk_luke}
\bz
{\cal S}\equiv \fr{(1+\slash v)}{2} [\slash S\gamma_5 -S_{0}^{*}]
\label{orb_sta}
\ez
with $S^{*}_{0}$ and $S_{\mu}$  the $0^+$ and $1^+$ fields respectively. 
The couplings $g_1$, $g_2$, $f_1$ and $f_2$ are unknown. In order to
obtain a conservative estimate of the widths we take the values
$g_1=g_2=f_1=f_2=0.50$. This choice
will probably result in an overestimate
of the widths given it does not respect the condition (\ref{sum_rule}).
Finally, we account for the the softening of the vertices with 
momentum by the factor \cite{ehq2}
\bz
e^{-p_{\pi}^2/\kappa^2} \fr{1}{(1+p_{\pi}^2/m_{\rho}^2)}
\label{mom_sup}
\ez
where $\kappa$ is a typical hadronic scale ${\cal O}(1)$~GeV.
This results in 
\bz
\ba{cc}
\G_{1}^{D}\simeq (0.25-0.35){\rm ~GeV}\;\;\; ; & \G_{2}^{D}\simeq (0.40-0.50) 
{\rm ~GeV} \\
\G_{1}^{B}\simeq (0.30-0.40) {\rm ~GeV}\;\;\; ; & \G_{2}^{B}\simeq
(0.40-0.50) {\rm ~GeV} 
\ea
\label{widths}
\ez
The resulting prediction for the pion momentum distribution in 
$B^0\to \pi^-\ell^+\nu$ is given by the dashed line in Fig.~4, 
together with the narrow width approximation result. 
The reason for the effect being small lies on the fact that,  
when the widths are incorporated into the fit of the model to the 
$D^0\to\pi^-\ell^+\nu$ branching ratio, the heavy mass independent parameter
$a_1/\gamma$ must be slightly modified to upset the width effect. 
Thus the same type of cancellation takes place in the $B\to \pi\ell\nu$
case. The procedure to obtain a prediction for the 
$B\to \pi$ transition is built in a way that is  nearly independent 
of the widths.  The model dependence on the cutoff $\L$ in the 
continuum integral is also marginally small. This is partly because 
of the fitting to the charm meson decay data, but mostly because 
the continuum plays a small role overall. 
Finally, we must address violation to the heavy mass independence of 
the ratios $a_i/\gamma$. The heavy mass dependence in these couplings
can be written as 
\by
a_i &\simeq & m_{H}^{3/2}\left( 1 +\fr{\alpha_i }{m_H} + ... \right) \nn\\
\gamma &\simeq & m_{H}^{3/2}\left( 1 +\fr{\beta }{m_H} + ... \right)\nn
\label{scal}
\ey
Implying 
\bz
\fr{a_i}{\gamma}\simeq N\left( 1+\fr{(\alpha_i-\beta)}{m_H} + ...\right)
\label{scal_frac}
\ez
with $N$ independent of $m_H$. 
Therefore potentially large $1/m_H$ corrections partially cancel in 
$a_i/\gamma$ and the use of the flavor HQS in the branch cut
is likely to be a very good approximation. 

In Fig.~5 we show the pion momentum distribution for the 
$D^0\to\pi^-\ell^+\nu$ decay. The solid line represents
the CDM prediction, whereas the dashed line corresponds to the 
chiral pole term and the dotted line is the BSW model. 
The CDM prediction agrees well with both, the first term 
in the dispersion relations and the BSW model. The maximum
pion momentum is not large enough to cause a disagreement, 
which arises at $p_{\pi}\gta 1$~GeV and to which only the CDM 
model is sensitive. 

As stated above, our results depend on the external parameters
$g$, $f_D$, $f_B$ and $Br^{D\to\pi}$, the $D^0\to\pi\ell^+\nu$  
branching ratio. In Table~I we present the results for the 
$B^0\to\pi^-\ell^+\nu$
branching ratio in units of $|V_{ub}|^2$ for a particular
choice of these external parameters 
and for two sets of masses for the 
resonances, as well as for the finite width case. The set of radial 
excitation masses we call Set~1 is the one presented in the text, 
whereas Set~2 corresponds to a shift of $+100$~MeV in the masses. 
As it can be seen, the result is rather stable under these type of 
modifications.
 In Table~2 we show results in the narrow width 
approximation for $f_D=0.24$~GeV and for various values of 
$g$ and $f_B$. As expected, the predicted branching ratio 
is rather sensitive to the values of these two external 
parameters, which hopefully will be determined in the 
near future by experiment and/or lattice calculations. 

Although the normalization of the model prediction for $B^0\to\pi^-\ell^+\nu$
depends on  poorly known or still unmeasured quantities, one 
characteristic feature of the model that emerges independently
of these is 
the shape of the momentum distribution. As is seen in Figs.~3~and~4, 
at pion momenta above $\approx 1$~GeV the CDM model predicts that the 
distribution flattens out, as opposed to the characteristic 
almost-linear growth of the pure $B^*$-pole behavior. 
In particular, the CDM prediction for the shape at maximum recoil
momentum ($q^2=0$) is consistent with the {\em shape} of the 
perturbative form-factor in the Brodsky-Lepage formalism. 
Considering that the CDM shape is a result of imposing the perturbative
behavior in the asymptotic region, e.g. for large $-q^2$, this coincidence
is interesting and deserves to be studied further.  
We expect 
the shape of the pion momentum distribution to be greatly constrained
by the future CLEO data \cite{law}. 
This will allow to discriminate between the CDM approach and the pure 
$B^*$-pole model. 
From Table~1 it can be seen that the intrinsic
uncertainty of the model is small. This is due to the fact that each 
change in a model parameter leads to a new value of $a_1/\gamma$ to 
keep a good fit of $Br^{D\to\pi}$. This procedure, built in the model, 
makes the predictions for $B^0\to\pi^-\ell^+\nu$ very stable. 
However, 
an extraction of $|V_{ub}|$ from the CLEO data is not possible at this 
point given the large uncertainties in external parameters, namely 
$g$, $f_B$ and $Br^{D\to\pi}$.

\section{Conclusions}
We have presented an approach to semileptonic form-factors that 
bridges the gap between low and high pion momentum recoil. 
The dispersion relation described in Section~2 
is a suitable framework to identify the various contributions. 
Furthermore, the imposition of model-independent theoretical constraints
leads to either complete calculations or an important reduction in
model dependence in calculating the various pieces entering in these
transitions. 

First, chiral symmetry tells us that , for $p_\pi\lta 1$~GeV, the first term
in the dispersion relation (\ref{un_dr}) is a good approximation 
to $f_+(t)$, with $\gamma$ given by (\ref{chi_gam}). The singularity 
closest to this region of the $H\to\pi\ell\nu$ decay is the leading
term entering in $f_+(t)$ in CPTHH. The next contribution to (\ref{un_dr})
comes from the $H^{(*)}\pi$ continuum, which dominates the $Im[f_+(s)]$
from threshold (the branch point) up to a scale where individual
resonances start dominating.
The phase of $f_+(s)$ in this region is computed in 
CPTHH, which is a valid tool up to this energy scale. 
We conclude that the resulting contribution is not the chief modification
of the $H^*$-pole behavior. 

On the other hand, the radial excitations dominate the $Im[f_+(s)]$ 
above the scale $\Lambda$, and result in large deviations from the 
$H^*$-pole behavior at values of $p_{\pi}$ larger than $\simeq 1$~GeV. 
Their contributions to the branch cut obey a definite HQS scaling 
with $m_H$. This imposes a constraint that allows us to relate 
the $B\to\pi\ell\nu$ form-factor to that entering in the 
$D\to\pi\ell\nu$ mode for {\em all values of the pion momentum}.
This represents a considerable improvement over the simple
$m_H$  scaling of the form-factors, which only allows a prediction of 
the $B$ mode up to pion momenta smaller than $1$~GeV. 

Finally, requiring that the form-factor has the correct 
asymptotic behavior as dictated by perturbative QCD, 
we derive the convergence relation or sum rule of (\ref{con_sr}). 
This translates into an additional constraint to be obeyed
by the couplings of the radial excitations, as seen in (\ref{con_nwa}). 

The model presented in Section~4 has two active resonances in the 
branch cut, leaving only one free parameter, which is fixed by 
fitting to the observed
$Br^{D\to\pi}$. The resulting form-factor, as seen for instance 
in (\ref{fp_nwa}) in the narrow width approximation, presents several
interesting features. Firstly, it respects the constraints 
from chiral symmetry, HQS and perturbative QCD. 
Secondly, it is a realization of the 
intuitive idea that the $H^*$-pole behavior must be softened by an effective
suppression of the coupling $\gamma$. In our model this suppression
is provided by the radial excitations in (\ref{fp_nwa}). 
This can be readily seen if we write the first term in (\ref{fp_nwa})
as
\bz
\fr{\gamma}{m_{H^*}^2-t}\left(\fr{M_{i}^2-m_{H^*}^2}{M_{i}^2-t}\right)
\simeq \fr{\gamma}{m_{H^*}^2-t}\left(\fr{1}{1+v.p_{\pi}/\Delta_i}\right)
\label{int_sup}
\ez
where $\Delta_i\equiv M_i-m_H$ is the gap between the ground state 
and the {\em i}-th excitation, and provides the scale of suppression 
 which in this case is of about $(0.8-1)$~GeV. Therefore the resonances
in the cut give the scale for the effective suppression of the $H^*$-pole 
behavior at large pion recoil. The model presents a natural explanation
for the softening of $\gamma$ \cite{bd_pole,iw_pole}. It also 
provides us with a simple rule for the validity of pole dominance
in semileptonic transitions in general: nearest pole dominance
is a good approximation for values of the hadronic recoil energy 
small compared to the gap between the first excitations
of the pole and the ground state.  This is the reason why 
pole dominance is a good approximation in $D$ semileptonic 
decays as well as in $B\to D^{(*)}\ell\nu$ decays, but not 
in $B\to\pi\ell\nu$, where large portions of the rate come from 
$v.p_{\pi}>\Delta_i$. 

As discussed in Section~4, the intrinsic uncertainty of the 
model is small. This implies that a determination of $|V_{ub}|$ 
from the measurement of the $B\to\pi\ell\nu$ branching ratio
will only be limited by the experimental precision 
with which $Br(D\to\pi\ell\nu)$, $g$, $f_D$ and $f_B$ are determined. 

The shape of the pion momentum distribution is a very distinctive 
feature of the model, as can be appreciated in Figs.~3~and~4. 
The comparison with the $B^*$-pole as normalized in CPTHH 
confirms that pole dominance is a good approximation for
$p_{\pi}\lta 1$~GeV. However, at larger values of $p_{\pi}$
the model deviates largely from the characteristic $B^*$-pole
distribution. For the same normalization at low $p_{\pi}$, the 
branching ratio is about a factor of two smaller than if the 
pure $B^*$-pole is assumed. The flat distribution at large pion momentum
is a consequence of the cancellations that result when 
imposing perturbative QCD as the asymptotic behavior. Interestingly, 
this shape resembles the one that is obtained by computing the 
form-factor directly in perturbative QCD for exclusive processes, 
characteristic of a dipole fall-off with 
$t$ \cite{gb_tes}, 
rather than a monopole. 
Although, as we argued in Section~3.3, the pQCD formalism is not 
expected to give the right answer, this coincidence in shape is 
an important point to bear in mind in an attempt to fully match 
the non-perturbative and perturbative regimes. 
Present and  
future CLEO measurements of the pion momentum distribution will
test this aspect of the model, independently of the values of external 
parameters. 

Another important test of this approach and the natural next step, is the 
prediction of $B\to\rho\ell\nu$. In this case the constraints from 
chiral symmetry are not so clear. However, the HQS and perturbative 
QCD are still present and largely define the behavior of the 
form-factors, as in the $B\to\pi\ell\nu$ case. 
Understanding the $B\to\rho\ell\nu$ to $B\to\pi\ell\nu$ ratio of branching
ratios, as well as the polarization in the vector mode will turn into a 
definite test of this approach and a major step towards the extraction 
of $|V_{ub}|$ from exclusive decays in the coming $B$ factory era. 

\begin{flushleft}
\Large\bf Acknowledgments
\end{flushleft}
The authors 
thank W. A. Bardeen, J. F. Donoghue, E. Eichten, L. Gibbons, 
C. Quigg, J. Simone and S. Willenbrock for useful discussions
 and suggestions.
G.B. acknowledges the support of the U.S. Department of Energy.
J.K. acknowledges support from the University of Massachusetts, Amherst.

\clearpage

\begin{table}
\begin{center}
{\large\bf Tables}
\end{center}
\vskip 0.3in
\centering
\begin{tabular}{|c|c|}
\hline
 &$Br(B^0\to\pi^-\ell^+\nu)/|V_{ub}|^2$ 
\\ \hline\hline
NWA~(Set~1)&14.00\\ \hline
NWA~(Set~2) &14.40\\ \hline
FW~(Set~1) &13.40\\ 
\hline
\end{tabular}
\vskip 0.10in
\caption{CDM prediction of the $Br(B^0\to\pi^-\ell^+\nu)$ for 
$g=0.50$, $f_D=0.24$~GeV and $f_B=0.15$~GeV. Set~1 corresponds to the 
spectrum of radial excitations of (30) , whereas Set~2
is obtained by a common shift of $+100$~MeV. The two first results
are in the narrow width approximation. The third one includes the effects
of the widths as discussed in Section~4, with widths given by 
(37).}
\end{table}
\vskip0.3in
\begin{table}
\centering
\begin{tabular}{|c|c|c|}
\hline
 &\multicolumn{2}{|c|}{$Br(B^0\to\pi^-\ell^+\nu)/|V_{ub}|^2$} \\
\cline{2-3}
$g$&\multicolumn{1}{|c|}{$f_B=0.15$~GeV}&\multicolumn{1}{|c|}{$f_B=0.20$~GeV} 
\\ \hline\hline
$0.30$&$19.50$&$35.00$\\ \hline
$0.50$&$14.00$&$25.00$\\ \hline
$0.70$&$7.30$&$14.00$\\ 
\hline
\end{tabular}
\vskip 0.1in
\caption{CDM prediction of the $Br(B^0\to\pi^-\ell^+\nu)$ in 
the narrow width approximation, for $f_D=0.24$~GeV
and with the Set~1 of radial excitation masses. The values chosen
for $g$ cover the allowed region. The results are shown for two
typical values of $f_B$. }
\end{table}

\clearpage
\vskip 0.1in
\noindent
\begin{center}
{\large\bf Figure Captions}
\end{center}
\vskip .5in

\noindent
Figure~1: Contributions to the imaginary part of 
the $H\to\pi\ell\nu$ form-factor $f_+(t)$.
The squares represent the action of the weak current and the circles 
imply strong interactions. The dashed lines are pions and the solid
lines are pseudoscalar heavy mesons $H$ unless indicated otherwise.
\vskip .2in
\noindent
Figure~2: The form-factor $f_+(t)$ as a function of the momentum 
transfer $t$. The solid line is the CDM prediction for 
$g=0.5$, $f_D=0.24$~GeV, $f_B=0.15$~GeV and $Br^{D\to\pi}=3.9\times 10^{-3}$. 
The dashed line is the $B^*$-pole prediction as normalized
by CPTHH 
in (\ref{chi_gam})
(chiral pole) and corresponds to the first term in the dispersion
relation (\ref{un_dr}). The dotted line is the BSW model 
prediction of Ref.~\cite{bsw}. 
\vskip .2in
\noindent
Figure~3:  The pion momentum distribution in units of $|V_{ub}|^2$ as
a function of the pion momentum. The caption is the same as in 
Fig.~2. 
\vskip .2in
\noindent
Figure~4: The pion momentum distribution in units of $|V_{ub}|^2$ as
a function of the pion momentum. The solid line is the CDM 
prediction in the narrow width approximation. The dashed line is 
the result taking into account finite widths, as given by (37). 
\vskip .2in
\noindent
Figure~5: The pion momentum distribution in $D^0\to\pi^-\ell^+\nu$ as
a function of the pion momentum. The solid line is the CDM fit to 
$Br^{D\to\pi}=3.9\times 10^{-3}$ for $g=0.5$ and  $f_D=0.24$~GeV. 
Also shown are the chiral pole (dashed line) and the BSW prediction
(dotted  line). 

\newpage
\begin{figure}
\vspace{15cm}
\includegraphics{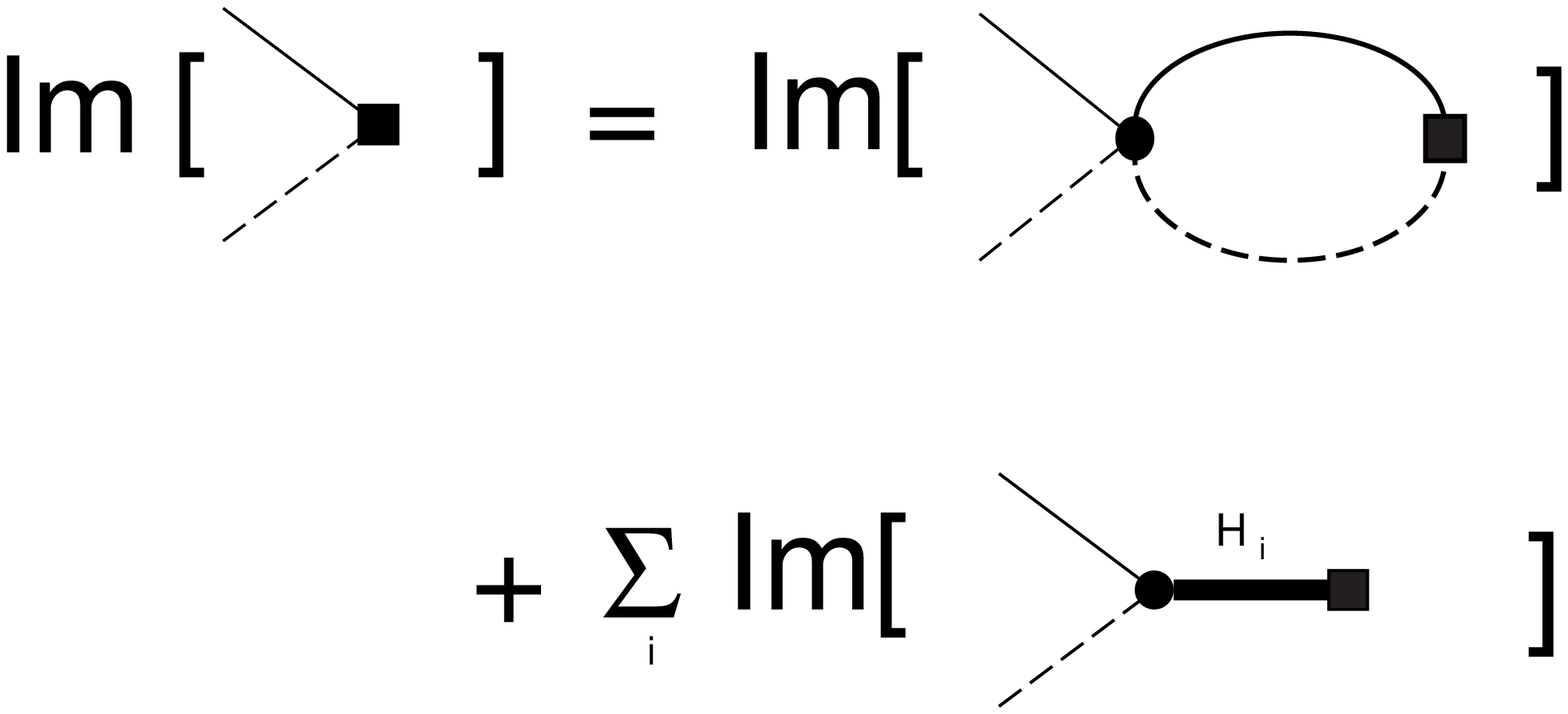}
\begin{center}
{\bf Figure 1}
\end{center}
\label{fig1}
\end{figure}

\newpage
\begin{figure}
\vspace{15cm}
\includegraphics{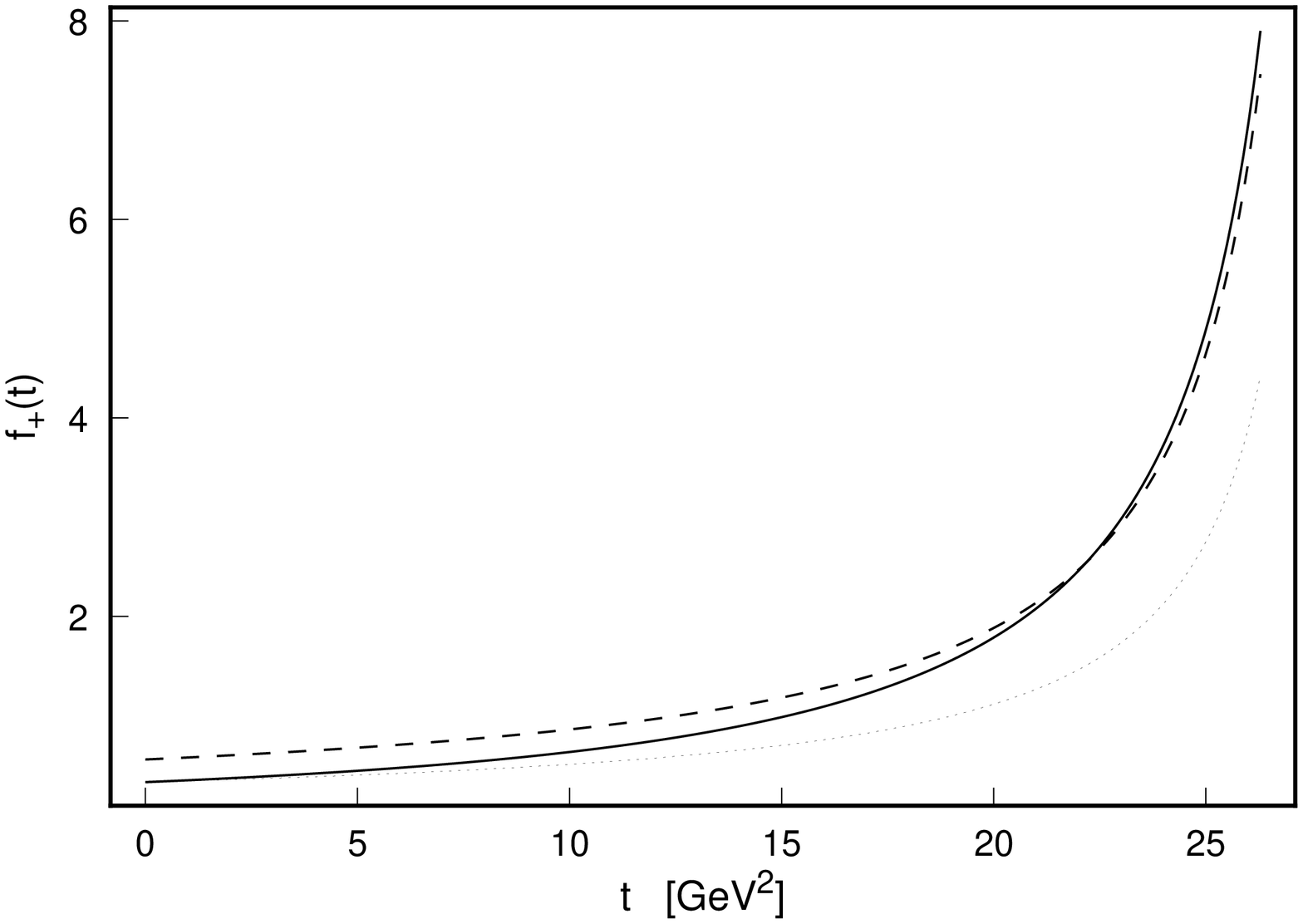}
\begin{center}
{\bf Figure 2}
\end{center}
\label{fig2}
\end{figure}

\newpage
\begin{figure}
\vspace{15cm}
\includegraphics{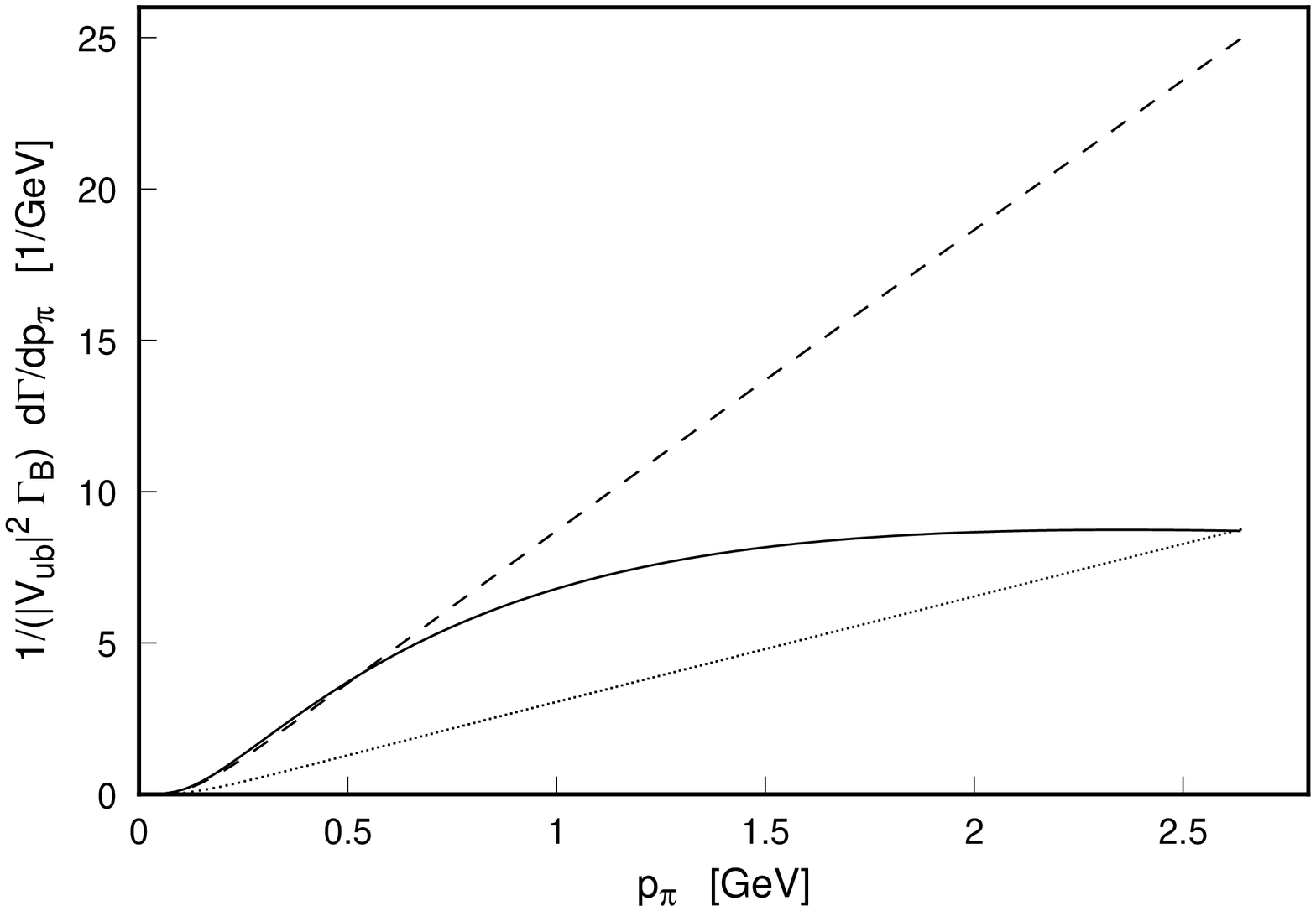}
\begin{center}
{\bf Figure 3}
\end{center}
\label{fig3}
\end{figure}

\newpage
\begin{figure}
\vspace{15cm}
\includegraphics{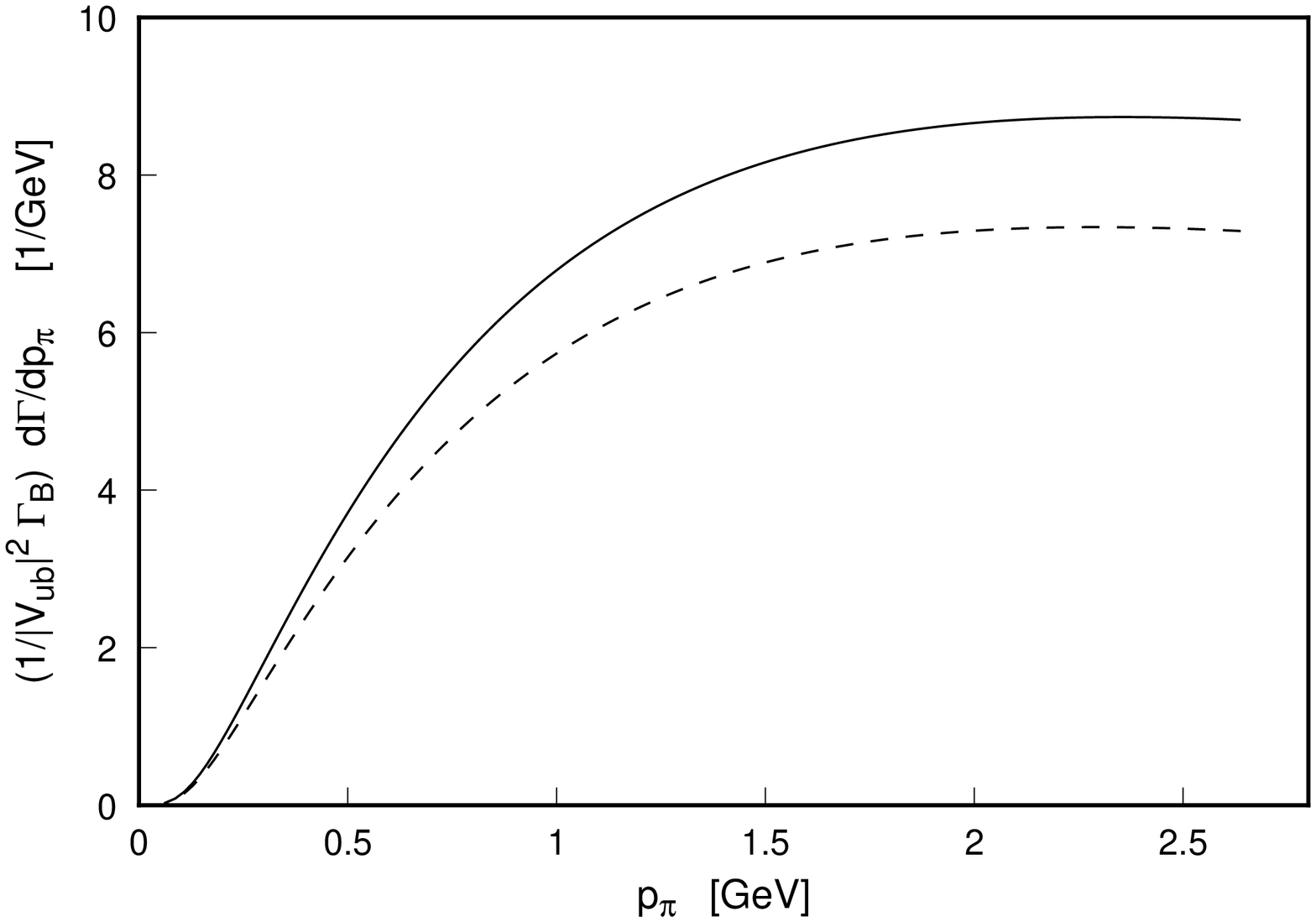}
\begin{center}
{\bf Figure 4}
\end{center}
\label{fig4}
\end{figure}

\newpage
\begin{figure}
\vspace{15cm}
\includegraphics{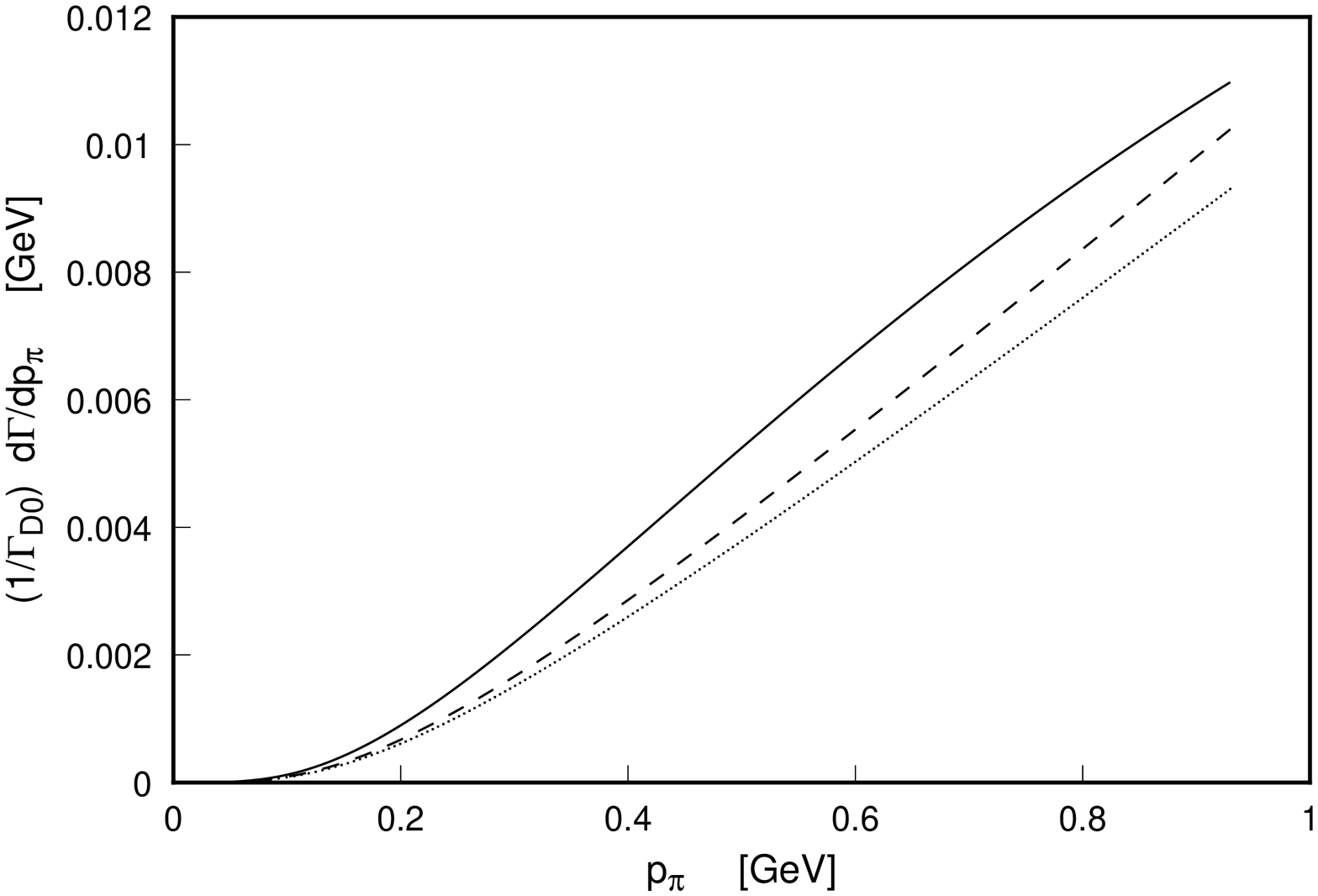}
\begin{center}
{\bf Figure 5}
\end{center}
\label{fig5}
\end{figure}


\begin{thebibliography}{99}

\bibitem{hqet} N. Isgur and M. B. Wise, Phys. Lett. {\bf B232}, 113 
(1989); {\it ibid} {\bf B237}, 527 (1990). For a review see M. Neubert, 
Phys. Rep. {\bf 245}, 259 (1994).   

\bibitem{hq_exclusive}M. Luke, Phys. Lett. {\bf B252}, 447 (1990); 
M. Neubert, Phys. Lett. {\bf B341}, 367 (1995). 

\bibitem{hq_inclusive}I. I. Bigi, M. Shifman, N. G. Uraltsev
and A. Vainshtein, Phys. Rev. Lett. {\bf 71}, 496 (1993); 
A. Manohar and M. B. Wise, Phys. Rev. {\bf D49}, 1310 (1994); 
M. Luke and M. Savage, Phys. lett. {\bf B321}, 88 (1994). 

\bibitem {wise} M.B. Wise, Phys.\ Rev.\ {\bf D45}, 2188 (1992).

\bibitem {bd} G. Burdman and J.F. Donoghue, Phys.\ Lett.\ {\bf B280},
287 (1992).

\bibitem{diff_bgl}Our approach is entirely different from that of 
C. Boyd, B. Grinstein and R. Lebed, Phys. Rev. Lett. {\bf 74}, 4603 
(1995). In this reference, bounds on the form-factors are
derived by computing a two-point function in pQCD at $q^2=16{\rm~GeV}^2$. 
However, at these values of $q^2$, large corrections are likely to 
exist that would affect the bounds. On the other hand, computing
at large {\em space-like} values of $q^2$, where pQCD should be trusted 
the most in the computation of the vacuum polarization involved,  
considerably loosens the bounds. 

\bibitem{cleo_bpi}R. Ammar {\em et al.}, the CLEO collaboration, 
CLEO CONF 95 09, EPS0165. 

\bibitem{ehq}E. Eichten, C. T. Hill and C. Quigg, FERMILAB-CONF-94/117-T, 
published in the Proceeding of the CHARM2000 Workshop, Fermilab, June 1994.

\bibitem{ehq2}E. Eichten, C. T. Hill and C. Quigg, Phys. Rev. Lett. 
{\bf 71}, 4116 (1993); and FERMILAB-CONF-94/118-T, published in 
the Proceeding of the CHARM2000 Workshop, Fermilab, June 1994. 

\bibitem{falk_luke}A. Falk and M. Luke, Phys. Lett. {\bf B292}, 
119 (1992). 

\bibitem{pqcd}G. P. Lepage and S. J. Brodsky, Phys. Lett. {\bf B87}, 959
(1979) and Phys. Rev. {\bf D22}, 2157 (1980). 
The first application to $B$ decays is found in A. Szczepaniak, E. M. Henley
and S. J. Brodsky, Phys. Lett. {\bf B243}, 287 (1990). 

\bibitem{sum_rule}C. A. Dominguez and N. Paver, Z. Phys. {\bf C41}, 
217 (1988); N. Paver and Riazuddin, Phys. Lett. {\bf B320}, 364 (1994)
and C. Chow and D. Pirjol, CLNS 95/1377. 

\bibitem{pdg} Particle Data Group, Phys. \ Rev. \ {\bf D50}, 1173 (1994).

\bibitem{fd_exp}D. Gibaut {\em et al.}, the CLEO collaboration, 
CLEO CONF 95-22, EPS0184. 

\bibitem{latt}R. M. Baxter {\em et al.}, the UKQCD collaboration, 
Phys. Rev. {\bf D49}, 1594 (1994); 
C. W. Bernard, J. N. Labrenz and A. Soni, Phys. Rev. 
{\bf D49}, 2536 (1994). For an update see C. Allton, Rome-95/1114, 
to be published in the proceedings of {\em Lattice '95}, Melbourne, 
Australia, July 1995. 

\bibitem{accmor}S. Barlag {\em et al.}, the  ACCMOR collaboration, 
Phys. Lett. {\bf B278}, 480 (1992). 

\bibitem{boy_ros} J. F. Amundson {\em et al.}, Phys.\ Lett.\
{\bf B296}, 415 (1992).

\bibitem{cleo_dp}F. Butler {\em et al.}, the CLEO collaboration, 
CLNS 95/1324, CLEO 95-3. 

\bibitem{will_johns} W. E. Johns, private communication.

\bibitem{bsw}M. Wirbel, B. Stech and M. Bauer, Z. Phys. {\bf C29}, 637 (1985). 

\bibitem{bd_pole} G. Burdman and J.F. Donoghue,  Phys. \ Rev. \  Lett.  
{\bf 68}, 2887 (1992).

\bibitem{isgw2}N. Isgur and D. Scora, Phys. Rev. {\bf D52}, 2783 (1995). 

\bibitem{isgw}N. Isgur, D. Scora, B. Grinstein and M. B. Wise, 
Phys. Rev. {\bf D39}, 799 (1989). 

\bibitem{law}L. Gibbons, private communication. 

\bibitem{iw_pole} N. Isgur and M. B. Wise, Phys. Rev. {\bf D41}, 151 (1990). 

\bibitem{gb_tes}G. Burdman and J. F. Donoghue, Phys. Lett. {\bf B270}, 
55 (1991). For a more detailed discussion see 
G. Burdman, UMI-94-08259-mc, 1993, Ph.D. Thesis.

\end{thebibliography}
\end{document}